\documentclass[%
 reprint,
superscriptaddress,
 amsmath,amssymb,
 aps,
]{revtex4-2}

\usepackage{graphicx}% Include figure files
\usepackage{dcolumn}% Align table columns on decimal point
\usepackage{bm}% bold math
\usepackage{xcolor}

\begin{document}

%\preprint{APS/123-QED}

\title{Insulating charge transfer ferromagnetism}

\author{Yixin Zhang}
\affiliation{Max Planck Institute for Chemical Physics of Solids, 01187, Dresden, Germany}
\affiliation{School of Physics, Peking University, Beijing 100871, China}

%\author{Nikolai Peshcherenko}%
%\affiliation{Max Planck Institute for Chemical Physics of Solids, 01187, Dresden, Germany}

%\author{Cheng Xu}
%\affiliation{Department of Physics and Astronomy, University of Tennessee, Knoxville, Tennessee 37996, USA}

\author{Yang Zhang}
%\email{yangzhang@utk.edu}
\affiliation{Department of Physics and Astronomy, University of Tennessee, Knoxville, Tennessee 37996, USA}
\affiliation{Min H. Kao Department of Electrical Engineering and Computer Science, University of Tennessee, Knoxville, Tennessee 37996, USA}


\begin{abstract}
We propose a mechanism for insulating ferromagnetism in the honeycomb Hubbard model of semiconductor moir\'e superlattices. The ferromagnetism emerges at critical charge transfer regime, stabilizing the quantum anomalous Hall state without Hund's coupling. We further note the ferromagnetic exchange applies to general charge transfer systems when breaking particle-hole symmetry.
\end{abstract}

\maketitle

%\tableofcontents
\iffalse
\section{Introduction}
\fi
\textit{Introduction.} Magnetism has become a major research focus in strongly correlated systems, driven by the complex interactions and emergent phenomena that arise in these materials \cite{makSemiconductorMoireMaterials2022, andreiMarvelsMoireMaterials2021, kennesMoireHeterostructuresCondensedmatter2021}. 
The Mott-Hubbard model remains a foundational framework for exploring magnetic behaviors, as it effectively captures the essential physics of electron-electron repulsion and kinetic energy, leading to the interaction of spin degrees of freedom \cite{hubbard1963electron,chao1978canonical}. This model has been instrumental in analyzing both insulating phases, where localized spins may organize into frustrated or ordered magnetic structures \cite{lee2006doping, tangEvidenceFrustratedMagnetic2023}, and metallic phases, where magnetism emerges from the dynamics of itinerant electrons \cite{nagaoka1966ferromagnetism,mielke1993ferromagnetism,ciorciaroKineticMagnetismTriangular2023, davydovaItinerantSpinPolaron2023,seifertSpinPolaronsFerromagnetism2024}. %The versatility of the Mott-Hubbard model allows it to describe a wide range of magnetic phenomena, from spin liquids and antiferromagnetic order to mettalic ferromagnetism,  spin polaron and kinetic magnetism, offering critical insights into the behavior of strongly correlated materials.
Transition metal dichalcogenide (TMD) moir\'e systems have recently emerged as a promising platform for exploring Hubbard physics. Significant experimental progress has been achieved, revealing various magnetic phases and novel transport properties in these systems \cite{xuInterplayTopologyCorrelations2024, zengThermodynamicEvidenceFractional2023, parkObservationFractionallyQuantized2023, caiSignaturesFractionalQuantum2023, xuObservationIntegerFractional2023, luFractionalQuantumAnomalous2024}. At large moir\'e periods, these systems can be effectively described by a tight-binding model, as the Wannier functions become well-localized, and the kinetic hopping between moir\'e sites is generally weaker than effective moir\'e potential \cite{wuHubbardModelPhysics2018a,xuMaximallyLocalizedWannier2024}. By incorporating on-site Coulomb interactions, such models are commonly employed to investigate the underlying magnetic properties of TMD moiré systems \cite{davydovaItinerantSpinPolaron2023, seifertSpinPolaronsFerromagnetism2024, zhangPseudogapMetalMagnetization2023, wuHubbardModelPhysics2018a, makSemiconductorMoireMaterials2022}.

Comparing other Hubbard systems, TMD moir\'e materials are highly tunable, via electrostatic gating, metallic gate distance, twist angle, stain, and magnetic field. Specifically, if we apply a gating field to control sublattice potential difference, the system can be driven from Mott-Hubbard insulator to a charge transfer insulator \cite{zhangEffectiveHamiltonianSuperconducting1988,zaanenBandGapsElectronic1985,zhangMoirQuantumChemistry2020}. More recently, for bilayer \(\mathrm{MoTe_2}\) with twist angle \(\theta \sim 3^\circ\), an unexpected ferromagnetic phase (FM) has been observed around filling factor \(\nu_h=1\) under the intermediate gating field \cite{zengThermodynamicEvidenceFractional2023}. Such FM phase is absent in either zero or large vertical electric fields, indicating that it cannot be solely attributed to Hund's coupling. Previous research indicates that the doped Hubbard model can exhibit metallic ferromagnetism through ring exchange, with several rigorous theorems established for the ground state spin sector \cite{nagaokaFerromagnetismNarrowAlmost1966,tasakiExtensionNagaokaTheorem1989,tasaki1998nagaoka}. In contrast, studies addressing insulating states often concentrate on the square and Kagome lattices, where nearest-neighbor interactions play a pivotal role \cite{zhuSpinOrderingTwodimensional1993, pollmannKineticFerromagnetismKagome2008}. There is limited exploration of the honeycomb lattice \cite{tangEvidenceFrustratedMagnetic2023}, especially when the insulating gap is of charge transfer nature. 

In this work, we study the magnetic phase diagram in the honeycomb Hubbard model with tunable charge transfer between sublattice sites. At filling \(\nu=1\), we focus on the insulating regime where the magnetic interactions are driven by exchange processes. We discover that with tunable charge-transfer, the system exhibits ferromagnetic phase at intermediate gating fields and antiferromagnetic phase (AFM) at extreme large fields. The charge-transfer FM arises from kinetic exchange rather than Hund's coupling. We demonstrate such a result by combining perturbation theory with exact diagonalization (ED) and density matrix renormalization group (DMRG) calculations for ground state energies and spin-spin correlations. We then extend our study to generalized Kane-Mele-Hubbard model, finding that ferromagnetism persists with complex second-nearest neighbor hopping. Our work introduces a new mechanism for ferromagnetism in charge transfer insulator without Hund's coupling, potentially explaining the experimentally observed $\nu_h=1$ ferromagnetism in twisted MoTe$_2$ at finite gating field \cite{zengThermodynamicEvidenceFractional2023}.

%In this work, we discuss the magnetism in honeycomb Hubbard model, considering that the potential between layers is in critical charge transfer regime.  Specifically, we will discuss the magnetic phase in the insulating case leading by the exchange process at filling \(\nu=1\). Here, \(\nu\) is defined as the number of electrons divided by the number of unit cells. We discover that in the insulating case if the interlayer exchange is suitably large, the system will be ferromagentic in the intermediate gating field but antiferromagnetic on both small and large extreme of the gating field. The ferromagnetic phase we have obtained arise purely from kinetic exchange rather than Hund's coupling (also known as direct exchange). We give a perturbation calculation analysis, along with exact diagonalization (ED) and density matrix renormalization group (DMRG) calculations for ground state energy and their correlations. We further discuss the topological case, discovering that the FM phase will persist when the second nearest hopping becomes complex. This work provides a new mechanics of feromagnetism of intermediate sublattice potential difference without Hund's coupling, paving the way of understanding experiment-observed magnetism and quantized transport signal. 
\begin{figure*}
    \centering
    \includegraphics[width=1\linewidth]{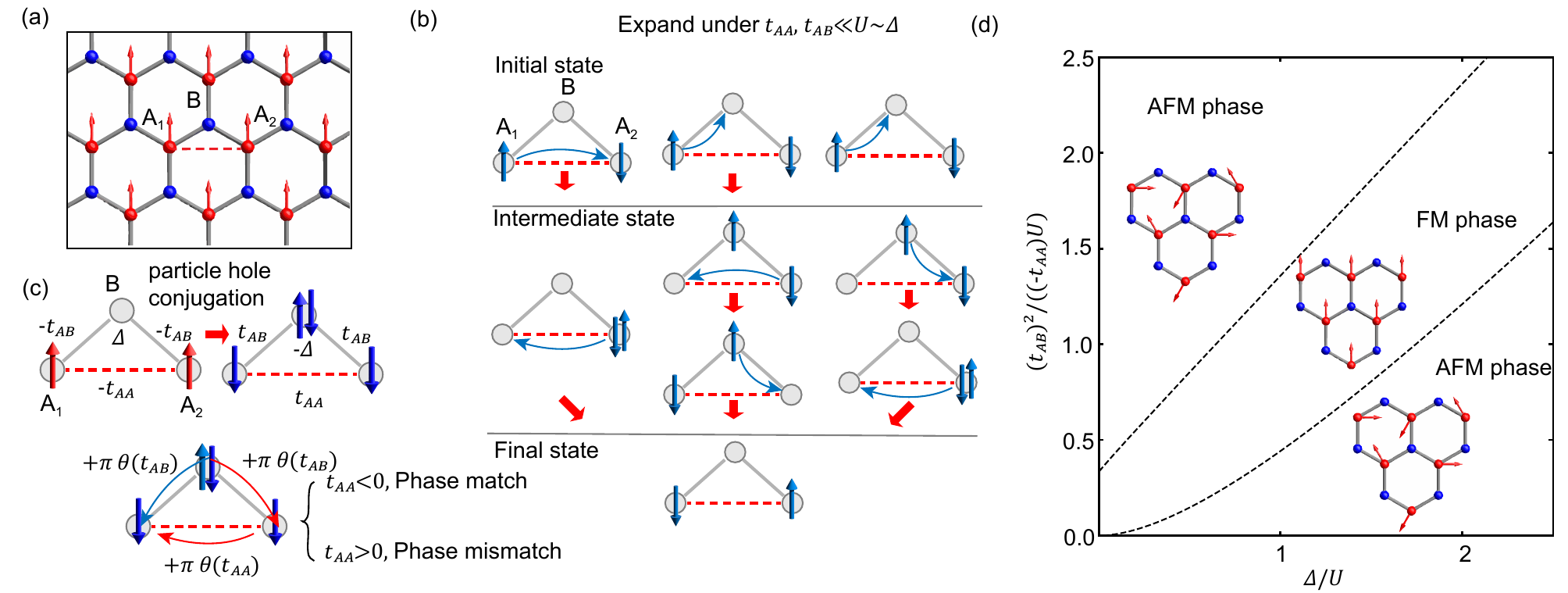}
    \caption{(a) Schematic diagram of a honeycomb lattice with two types of sites, A and B, illustrated in red and blue. The nearest neighbors are connected by thick gray tubes, while the second nearest neighbor is connected by red dashed lines. (b) Three exchange process for swapping two electrons spin on two A sites (cation site), by introducing an extra B site. (c) The schematic kinetic diagram for ferromagnetic ring exchange from \(t_{AA}<0\). \(\theta(x)\) stands for Heaviside step function. Two interfering paths are marked with blue and red arrows. (d) The phase diagram which is universal from the value of \(U\) and specific lattice, calculated from \(J>0\) in Eq.~\eqref{effective exchange full}. The phase diagram assumes \(U_A=U_B=U\).
    }
    \label{Fig1}
\end{figure*}

\iffalse
\section{Hubbard model description and AFM state at extreme case of \(\Delta\)}
\fi

\textit{Honeycomb Hubbard model and AFM state at large bias limit.} The starting point for our analysis is the bipartite canonical Hubbard model with on-site potential difference, featuring A and B sublattices of honeycomb network as shown in Fig.~\ref{Fig1}(a). The model Hamiltonian is written as:

\begin{align}
&\mathcal{H}=H_A+H_B+H_{AB}+U_A\sum_{i\in A}n_{i\uparrow}n_{i\downarrow}+U_B\sum_{i\in B}n_{i\uparrow}n_{i\downarrow}. \\
&H_{\alpha}=-\sum_{\langle i,j\rangle_{\alpha},\sigma} (t_{\alpha \alpha} c_{i\sigma}^{\dagger}c_{j\sigma}+\mathrm{H.c.})-\sum_{i\in\alpha}\frac{\tau_{\alpha}\Delta}{2}n_{i}, \\
&H_{AB}=-t_{AB}\sum_{\langle i,j\rangle,\sigma} c_{i\sigma}^{\dagger}c_{j\sigma}. \label{complete hamiltonian}
\end{align}
Here we define \(\alpha \in \{A, B\}\) with \(\tau_A = 1\) and \(\tau_B = -1\) and \(\sigma=\pm\) to denote the spin degree of freedom. We use \(\langle i,j\rangle_{\alpha}\) to denote nearest neighbor pairs within \(\alpha\) sublattice, and \(\langle i,j\rangle\) for nearest neighbor pairs between A and B sites. \(\Delta\) represents the potential difference between sublattices, while \(U_A\) and \(U_B\) are the on-site Coulomb potentials. For simplicity, we consider electron filling of the moir\'e band, where the electron filling factor is defined as \(\nu=n_{ele}/n_{cell}\). We further simplify the notation by setting \(U_A = U_B = U\), though the distinction will be reinstated in the explicit perturbation analysis. 

In the following, we investigate the corresponding phase diagram as a function of \(\Delta/U\). We start with a well-separated energy scale: \(|t_{AA}|, |t_{BB}| \ll U_A, U_B, \Delta\), where \(U\) and \(\Delta\) are assumed to be of the same order. Additionally, we assume \(|t_{AB}| \sim \sqrt{|t_{AA}| U} > |t_{AA}|\), which defines the parameter region where ferromagnetism can be clearly distinguished in the perturbation analysis. Numerical results studying wide range of parameters will be presented in the subsequent section.

At filling factor $\nu = 1$ in the strong coupling limit \(U \gg |t_{AB}|, |t_{AA}|\), having electron occupancy at B sites or forming double occupancy on A sites results in a significant increase in total energy. Consequently, the ground state, in the absence of kinetic hopping, consists of one electron localized on each A site (see Fig.~\ref{Fig1}(a)), with an extensive degeneracy from spin degrees of freedom. As the kinetic hopping becomes non-zero, this spin degeneracy is lifted, allowing for the development of an effective spin model \cite{chao1978canonical}.

Under large potential bias \(\Delta \gg U\), the model reduces to a triangular spin lattice model on A sublattice with antiferromagnetic exchange proportional to \(t_{AA}^2/U\). This leads to the well-known \(120^\circ\) AFM phase. On the other hand, for small \(\Delta\), particularly when it becomes comparable to \(t\), electrons delocalize on the B sublattice (the sublattice with higher energy) and higher-order terms such as superexchange interactions proportional to \(t_{AB}^4/(\Delta^2 U)\) favor antiferromagnetism. As \(\Delta\) decreases, these interactions can overcome possible ferromagnetic tendencies, leading to an AFM phase. It should be noted that if \(\Delta\) becomes the same order as \(t\), perturbation theory may no longer be valid, and higher-order terms must be considered. However, numerical analysis confirms a predominantly non-ferromagnetic ground state at \(\Delta=0\) for \(|t_{AB}| \sim \sqrt{|t_{AA}| U}\).

\iffalse
\section{Effective spin model and Ferromagnetic ring exchange}
\fi
\textit{Effective spin model and ferromagnetic ring exchange.} For intermediate gating fields \(\Delta/U \sim 1\), we develop an effective Heisenberg model using perturbation theory. We find that when \(t_{AA}/U\) is small, the phase determined from the model is governed by two dimensionless quantities \(-t_{AB}^2/(U t_{AA})\) and \(\Delta/U\), remaining independent of specific lattice structure when both quantities are \(\sim 1\). The resulting phase diagram is illustrated in Fig.~\ref{Fig1}(d).

The effective spin model is (derivation in Appendix~\ref{analytical calculation}):
\begin{equation}
    \mathcal{H}_{\text{effective}} = -J \sum_{\langle i,j \rangle_A} \boldsymbol{S}_i \cdot \boldsymbol{S}_j
\end{equation}
where \(S\) represents the on-site spin (excluding \(\hbar\)), and \(\langle i,j \rangle_A\) sums over all neighboring pairs within the A sites. The expression for \(J\) is given by:
\begin{eqnarray}
    J &=& - \frac{4t_{AA}^2}{U_A} - \frac{4t_{AB}^2 (t_{AA} U_A + 2t_{AA} \Delta)}{U_A \Delta^2}\nonumber\\ 
    &&- \frac{4t_{AB}^4(2U_A + U_B + 2\Delta)}{U_A \Delta^2 (U_B + 2\Delta)}\label{effective exchange full}
\end{eqnarray}

Two unique terms emerge beyond conventional exchange and superexchange interactions: \(t_{AB}^2 t_{AA}/(\Delta U)\) and \(t_{AB}^2 t_{AA}/\Delta^2\), arising from triangular loops within the bipartite lattice. The former represents a combination of kinetic exchange and superexchange. In such a process, an electron on A site hops to another A site through one exchange channel and returns via another (see Fig.~\ref{Fig1}(b) right). The latter corresponds to ring exchange (see Fig.~\ref{Fig1}(b) middle), sharing similarities with Nagaoka ferromagnetism \cite{nagaoka1966ferromagnetism,tasakiExtensionNagaokaTheorem1989, tasakiHubbardModelIntroduction1997, tasakiNagaokaFerromagnetismFlatband1998}. However, the crucial difference is that the Nagaoka ferromagnetism only appears in the metallic regime.

Given third-order kinetic exchange, the emergence of ferromagnetism requires \(t_{AA}<0\). This negative hopping allows different signs between channels with amplitude \(t_{AB}^2 /\Delta\) and \(t_{AA}\), making the contribution favoring ferromagnetism. This contrasts with single-channel hopping, which typically favors antiferromagnetism (see Fig.~\ref{Fig1}(b) left). For ring exchange, particle-hole conjugation transforms the system where two electrons are in three sites of the minimal triangle into the corresponding system with one excess electron hopping on a singly occupied background. %When the background spins are polarized, the hole binds to a B site, facilitated by constructive interference (phase match) as it encircles the triangular loop under \(t_{AA}<0\) as shown in Fig.~\ref{Fig1}(c)). 
When the background spins are polarized, the electron hopping from B site to A site is facilitated by constructive interference (phase match under \(t_{AA}<0\)) between two paths as shown in Fig.~\ref{Fig1}(c)). 

%\Red{The presence of odd-number hopping processes can be attributed to the breaking of particle-hole symmetry, making mechanics which electron or hole dop is required like Nagaoka ferromagnetism to emerge in insulating regime.}

\iffalse
\section{Real hopping calculation}
\fi
\textit{Real hopping calculation.} We now examine realistic parameters for the honeycomb lattice using \(U_A = U_B = 40 |t_{AA}|\), based on twisted bilayer \(\mathrm{MoTe_2}\) at \(3^\circ\) twist angle \cite{xuMaximallyLocalizedWannier2024}. We set \(t_{AA} = t_{BB}\) and also perform calculations with \(U_A = U_B = 500 |t_{AA}|\) at strong coupling limit for comparison, which we expect a better match for a smaller \(t_{AA}/U\) with perturbation analysis (see Appendix~\ref{large U real hopping}).

\begin{figure}
    \centering
    \includegraphics[width=1\linewidth]{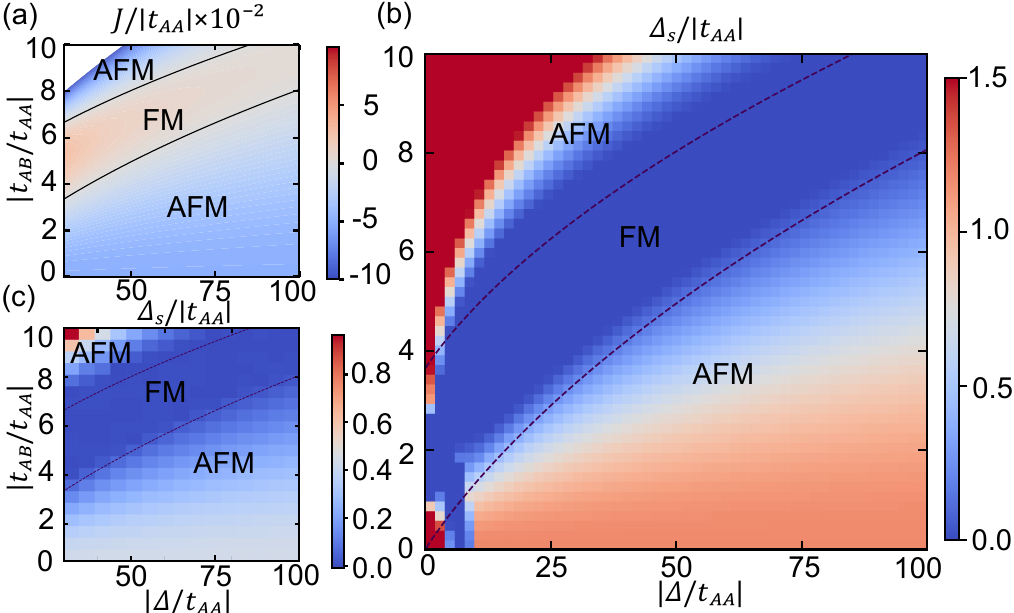}
    \caption{(a) Analytical phase diagram of the strength of \(J/|t_{AA}|\) as a function of \(t_{AB}\) and \(\Delta\). (b) The spin gap \(\Delta_s\) in a \(3\times3\) PBC  cluster using ED. A cutoff of 1.5 is imposed for clarity. (c) The spin gap \(\Delta_s\) in a \(3\times12\) cylinder using DMRG. In (b) and (c), the dashed line corresponds to the phase boundary in (a).}
    \label{Fig2}
\end{figure}

The strength of \(J\) with respect to \(t_{AB}\) and \(\Delta\), calculated by incorporating \(U_A = U_B = 40 |t_{AA}|\) into Eq.~\eqref{effective exchange full} is shown in Fig.~\ref{Fig2}(a). At zero temperature and within the scope of two-body interaction, the system shows ferromagnetic (FM) states with \(J > 0\) and antiferromagnetic (AFM) states with \(J < 0\), valid when \(t \ll \Delta\).

We verify this analytically derived phase boundary using exact diagonalization (ED) on PBC (periodic boundary condition) cluster with \(L_x=3\) and \(L_y = 3, 6\) and density matrix renormalization group (DMRG) methods on three-leg cylinder with \(L_y = 12\). The numerical results of the honeycomb Hubbard model align well with the effective spin model predictions, especially for large \(\Delta\) compared to \(t_{AA}\) and \(t_{AB}\). Periodic boundary conditions were applied in both directions for ED and in the x direction for DMRG. Note that we divide the total Hilbert space into \(N_\uparrow\) and \(N_\downarrow\) sectors using particle number \(N\) and total spin \(S_z\) conservation. Details of the calculations can be found in Appendix~\ref{Calculation Detail for Real Hopping}.

\iffalse
\subsection{The Spin Gap}
\fi
As the indicator for FM and AFM phases, we calculate spin gaps of spin flips using ED and DMRG on \(3 \times L_y\) clusters. For ED on a \(3 \times 3\) PBC clusters, we focused on the spin-4 gap, measuring the energy difference between the fully polarized state \(S_z = S_{\text{max}}=9/2\) (the fully polarized sector) and the \(S_z = 1/2\) state (containing the true ground state). %Alternatively we can say "Examing if such spin gap reach zero provides strictest  "
For DMRG on a \(3 \times 12\) cylinder, however, due to the large Hilbert space and nonzero occupation in $B$ sublattice, we calculated only the energy difference between the states with \(S_z = S_{\text{max}} - 1\) and \(S_z = S_{\text{max}} - 2\), commonly referred to as the spin-1 gap. Although spin gap of multiple spin flips better indicates AFM energy gain compared to the spin-1 gap \cite{davydovaItinerantSpinPolaron2023}, we show that the different spin gaps are qualitatively similar for system with $SU(2)$ symmetry, based on calculations on PBC \(3 \times 3\) clusters.

Fig.~\ref{Fig2}(b) shows the spin gap dependency on \(t_{AB}\) and \(\Delta\) for \(3\times 3\) cluster. The dashed line indicates the perturbation theory phase boundary prediction, which qualitatively agrees with the calculated spin gap. Better agreement is observed for larger \(U, \Delta\) (see appendix~\ref{large U real hopping}). Fig.~\ref{Fig2}(c) displays similar behavior for the 3 × 12 DMRG calculation, suggesting the magnetism is robust in the thermodynamic limit.

\iffalse
\subsection{The Spin-Spin correlation}
\fi

We also analyze spin-spin correlation across our parameter space, observing significant differences between regions where saturating field \(h = 0\) and \(h \neq 0\) in ED calculations. This approach identifies various AFM phases, including the \(120^\circ\) AFM and possible non-magnetic phases, and can be extended to cases with complex \(t_{AB}\) where total spin is not conserved. DMRG calculations show similar results, as detailed in Appendix~\ref{Extra Data for Spin Correlation}.

The spin-spin correlation function is defined as:
\begin{equation}
    \chi(i,j) = \langle \boldsymbol{S}_i \cdot \boldsymbol{S}_j \rangle
\end{equation}
where \(i,j\) label different sites, and \(S^\lambda_i = \frac{1}{2} c_{i, \sigma_1}^\dagger \tau_{\sigma_1 \sigma_2}^\lambda c_{i, \sigma_2}\), with \(\tau_i\) being the Pauli matrix. This \(\mathrm{SU}(2)\) invariant quantity is calculated in the ground state mainifold.

\begin{figure}
    \centering
    \includegraphics[width=1\linewidth]{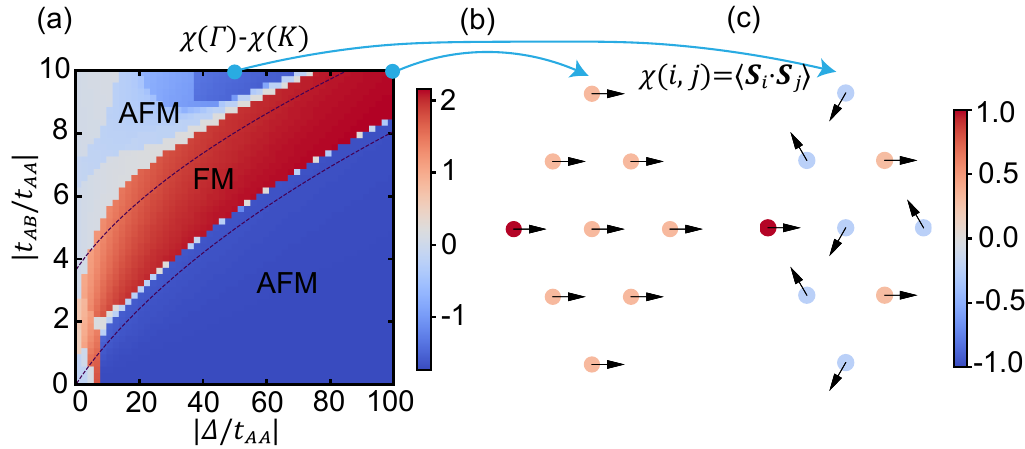}
    \caption{(a) The spin-spin correlation intensity difference between the \(\Gamma\) and \(K\) points in a \(3\times3\) PBC  cluster using ED. A larger value indicates greater concentration at the \(\Gamma\) point. (b, c) Normalized spin-spin correlation by self-correlation in real space at \(t_{AB} = 10 |t_{AA}|\), with (b) \(\Delta = 100 |t_{AA}|\) and (c) \(\Delta = 50 |t_{AA}|\). Arrows on sites are included for visual guidance.}
    \label{Fig3}
\end{figure}

Spin correlations in FM and AFM regions show expected phase distributions: fully polarized in FM and \(120^\circ\) AFM phase in AFM region (Fig.~\ref{Fig3}b, c; Appendix~\ref{Extra Data for Spin Correlation}). We identify those states from k-space correlation by defining the indicator:
\begin{equation}
    \chi_{\text{diff}}=\chi(\Gamma)-\chi(K),
\end{equation}
where:
\begin{equation}
    \chi(\boldsymbol{k}) = \sum_j \chi(i,j) e^{i \boldsymbol{k} (\boldsymbol{r}_i - \boldsymbol{r}_j)} = \langle \boldsymbol{S}(\boldsymbol{k}) \cdot \boldsymbol{S}(-\boldsymbol{k}) \rangle
\end{equation}
FM phases peak at the \(\Gamma\) point due to translation invariance, while \(120^\circ\) AFM phases peak at the \(K\) point. Fig.~\ref{Fig3}(a) shows results, with large \(\Gamma\) peak region overlapping zero spin gap regions, validating our method.

\begin{figure}
    \centering
    \includegraphics[width=1\linewidth]{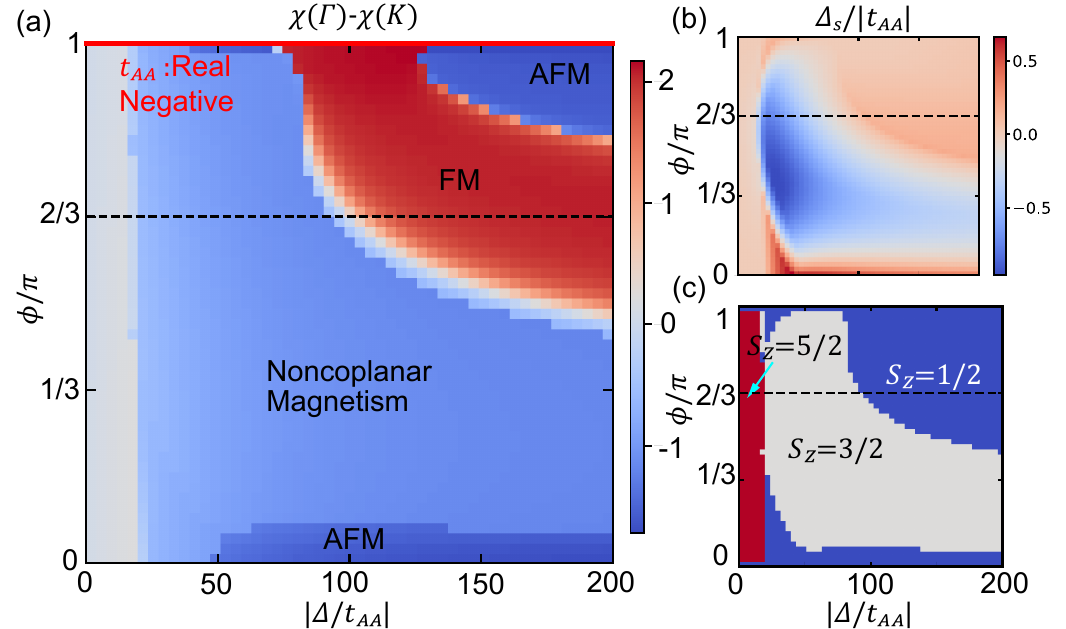}
    \caption{Calculation results for \(t_{AB} = 10 |t_{AA}|\), \(t_{AA} = |t_{AA}| e^{i \phi} \), \(U_A = U_B = 40 |t_{AA}|\), \(\Delta = 150 |t_{AA}|\) in a \(3\times3\) PBC cluster using ED. (a) The difference in spin-spin correlation between the \(\Gamma\) and \(K\) points. (b) The energy difference \(\Delta_s\) between \(S_z = 3/2\) and \(S_z = 1/2\), also known as the spin-one gap. (c) The \(S_z\) sector in which the ground state lies. Note that although the ferromagnetic state has a small total \(S_z\), it exhibits strong ferromagnetic correlation.}
    \label{Fig4}
\end{figure}

\iffalse
\section{Complex hopping calculation}
\fi
\textit{Complex $t_{AA}$.} We now investigate the charge-transfer ferromagnetism under complex second-nearest neighbor hopping, adjusting the Hamiltonian to the Kane-Mele-Hubbard model (see Appendix~\ref{Effective spin model for complex hopping} for effective spin model). With complex \(t_{AA} = |t_{AA}| e^{i \phi}\), spin-spin correlation serves as the more reliable phase indicator as SU(2) symmetry is broken. Fig.~\ref{Fig4}(a) shows that maximum correlation strength remains relatively unchanged with varying \(\phi\), indicating persistent ferromagnetism. The ferromagnetic region shifts towards larger \(\Delta\). A detailed explanation is provided in Appendix~\ref{Effective spin model for complex hopping}.

We analyzed z-direction spin texture by identifying the ground state spin sector (Fig.~\ref{Fig4}(c)) and the energy difference between \(S_z = 3/2\) and \(S_z = 1/2\) sectors (Fig.~\ref{Fig4}(b)). As shown from the figure, in the ferromagnetic region, there will be a small energy gain for having \(S_z = 1/2\), indicating in-plane ferromagnetism. The AFM state outside this boundary shows spin-unpolarized and weakly z-polarized regions, with the latter resembling canted magnetic states \cite{qiu2023interaction}. This likely results from the Dzyaloshinskii–Moriya interaction in the spin model \cite{yangFirstprinciplesCalculationsDzyaloshinskii2023}.

%The reduction in AFM strength and the shift of the FM region toward larger \(\Delta\) values suggest a pathway to achieving ferromagnetism under large potential differences, highlighting the interplay between topology and magnetic structure. This scenario presents an ideal platform for investigating quantum anomalous Hall (QAH) and fractional quantum anomalous Hall (FQAH) physics. It is conjectured that a similar physical mechanism may apply to the filling factor \(\nu = 2/3\). Although this is beyond the scope of this article, we briefly note that achieving such a state would require introducing nearest-neighbor interactions. The effective process shown in Fig.~\ref{Fig1}(b) would still persist, with the potential replaced by the effective interaction.

\iffalse
\section{Discussion}
\fi
\textit{Discussion.} Our work identifies a new type of ferromagnetism in moir\'e superlattice driven by kinetic exchange, without the needs of Hund's coupling. With analytical perturbation analysis, ED and DMRG calculations, we present a universal phase diagram dependent as a function of \(\Delta/U\) and \(t_{AB}^2/(t_{AA} \Delta)\), showing ferromagnetism at intermediate \(\Delta\) and antiferromagnetism at small and large \(\Delta\). Our charge-transfer exchange can induce ferromagnetism by adjusting gating fields, as long as triangle loops connect two spins in the lower-energy sublattice with an unoccupied site in the higher-energy sublattice.

The presence of a gating field breaks particle-hole symmetry, allowing terms in the effective exchange that favor ferromagnetism, characterized by the product of three hopping strengths in the numerator (ie. \(t_{AB}^2 t_{AA}/\Delta^2, t_{AB}^2 t_{AA}/(\Delta U)\)). The $t^3$ exchange terms change sign under particle-hole conjugation, which changes the hopping strengths \(t_{ij} \to -t_{ij}\) and the onsite potential \(V \to -V\). This contrasts with the half-filled bipartite lattice with equal onsite potential, which preserves particle-hole symmetry. There, only exchange terms proportional to the product of an even number of hopping strengths exist, favoring antiferromagnetism when all hopping are considered real, proved by Lieb \cite{liebTwoTheoremsHubbard1989}. This sheds light on a general paradigm for ferromagnetism in insulating cases: breaking particle-hole symmetry enables a product of an odd number of hopping strengths in the effective exchange, providing a trend toward ferromagnetism when the hopping has a specific sign.

The charge transfer model can be realized in several realistic material systems, particularly R-stack moiré bilayers where minibands originate from \(\Gamma\) or \(K\) valley. In the former case, the formalism of real \(t_{AA}\) applies whereas in the latter case consideration of complex hopping is needed. The detailed discussions are presented in Appendix~\ref{Connected to real material calculation}. In particular, charge-transfer ferromagnetism observed in R-stacked twisted bilayer \(\mathrm{MoTe_2}\) around \(\nu_h = -1\) under intermediate gate fields,  along with antiferromagnetism at small and large gating, can be directly explained by our mechanism, rather than the dominant Hund's coupling at large twist angle regime. Our mechanism can potential apply to heterobilayer MoTe$_2$/WSe$_2$ 
\cite{li2021quantum,tao2024valley}.

%This potentially explains ferromagnetism observed in twisted bilayer \(\mathrm{MoTe_2}\) around \(\nu = -1\) under intermediate gate fields. The essential physics can be captured by a Mott-Hubbard model with negative \(t_{AA}\) (Appendix~\ref{Connected to real material calculation}). Such mechanism becomes more significant as electron localization increases with gating, ruling out Hund's coupling as a cause since no magnetism is observed even in case of a small gating field. Our mechanism, emerging from bulk properties, remains robust for doping much less than \(t_{AA}/U\). Different charge transfer processes emerge upon further doping.

While our calculations focus on a honeycomb lattice-based model, the theory extends to other systems with triangular loops across two sublattice sites, such as superposed chiral \(\pi\)-flux models on checkerboard lattice \cite{regnaultFractionalChernInsulator2011, neupertFractionalQuantumHall2011}. The charge-transfer induced ferromagnetism provides a alternate route for stabilizing quantum anomalous Hall phases at strong coupling limit. Our theory can also be extended to systems with slight doping below the order of \(\frac{t_{AA}}{U}\), where the kinetic energy of the doped electrons does not exceed the overall FM or AFM gain. We will present the detailed study of the metallic charge transfer ferromagnetism from kinetic hopping at strong coupling limit with $t\ll U\sim \Delta$ in a separate work \cite{kineticFM}. 

%This work opens avenues for exploring magnetic aspects of integer and fractional topological phases observed experimentally, with potential extensions to compute Chern numbers and include interaction terms like \(\sum V_{ij} n_i n_j\).

\textbf{Acknowledgement.} We acknowledge the collaboration with Nikolai Peshcherenko, Ning Mao, and Shengwei Jiang, and helpful discussions with Cristian Batista and Shu Zhang. Yang Zhang is supported by the start up grant at University of Tennessee Knoxville and Max-Planck Partner lab from Max Planck Institute Chemical Physics of Solids.

\bibliography{apssamp}% Produces the bibliography via BibTeX.

\clearpage % Start appendix on a new page
\appendix
\onecolumngrid
\section*{Supplementary material for: Insulating charge transfer ferromagnetism}

% Reset counters and redefine numbering if necessary
\setcounter{section}{0}
\setcounter{figure}{0}
\setcounter{equation}{0}
\renewcommand{\thefigure}{S\arabic{figure}}

\section{\label{analytical calculation}From Hubbard model to effective spin model}
\begin{widetext}
To understand the emergence of ferromagnetism through perturbation analysis, we involve virtual hopping processes. By examining the processes that connect low-lying states, we can determine the energy differences between various spin configurations and derive the effective spin model. \cite{chao1978canonical, maoWannierFunctionsMinimal2023}. 

We focus on virtual processes involving fewer than four hoppings and only include two spins. These processes are confined to three atoms forming a connected triangle, including two sites in the A sublattice where spins are initially localized and one site in the B sublattice. In Fig.~\ref{Fig1}(b) of the main text we list several ways that can exchange spins on two lattice sites. Here, we assume that one spin is up and the other is down, since any state with both spins up or down can be converted to this scenario by applying \(S_1^- + S_2^-\) or \(S_1^+ + S_2^+\) without altering the interaction energy. The corresponding tight-binding model is:
\begin{align}
H &= \epsilon_a \sum_{\sigma} \left( c_{A_1 \sigma}^\dagger c_{A_1 \sigma} + c_{A_2 \sigma}^\dagger c_{A_2 \sigma} \right) + (\epsilon_a + \Delta) \sum_{\sigma} c_{B \sigma}^\dagger c_{B \sigma} \\
&\quad - t_{AB} \sum_{\sigma} \left( c_{A_1 \sigma}^\dagger c_{B \sigma} + c_{A_2 \sigma}^\dagger c_{B \sigma} + \text{h.c.} \right) \\
&\quad - t_{AA} \sum_{\sigma} \left( c_{A_2 \sigma}^\dagger c_{A_1 \sigma} + \text{h.c.} \right) \\
&\quad + U_A \sum_{i=A_1,A_2} n_{A_i\downarrow} n_{A_i\uparrow} + U_B n_{B\downarrow} n_{B\uparrow}.
\end{align}
We expand under the basis:
\begin{equation}    
\left\{
c_{A_1\downarrow}^\dagger c_{A_2\uparrow}^\dagger\left.| 0 \right\rangle,
c_{A_1\uparrow}^\dagger c_{A_2\downarrow}^\dagger\left.| 0 \right\rangle,
c_{A_2\uparrow}^\dagger c_{B\downarrow}^\dagger\left.| 0 \right\rangle,
c_{A_1\downarrow}^\dagger c_{B\uparrow}^\dagger\left.| 0 \right\rangle,
c_{A_2\downarrow}^\dagger c_{B\uparrow}^\dagger\left.| 0 \right\rangle,
c_{A_1\uparrow}^\dagger c_{B\downarrow}^\dagger\left.| 0 \right\rangle,
c_{A_2\uparrow}^\dagger c_{A_2\downarrow}^\dagger\left.| 0 \right\rangle,
c_{A_1\uparrow}^\dagger c_{A_1\downarrow}^\dagger\left.| 0 \right\rangle,
c_{B\uparrow}^\dagger c_{B\downarrow}^\dagger\left.| 0 \right\rangle
\right\}
\end{equation}
The corresponding Hamiltonian matrix in this basis is:
\begin{equation}
    \mathcal{H}=\begin{pmatrix}
2 \epsilon_a & 0 & t_{AB} & -t_{AB} & 0 & 0 & t_{AA} & t_{AA} & 0 \\
0 & 2 \epsilon_a & 0 & 0 & t_{AB} & -t_{AB} & -t_{AA} & -t_{AA} & 0 \\
t_{AB} & 0 & \Delta + 2 \epsilon_a & 0 & 0 & -t_{AA} & -t_{AB} & 0 & -t_{AB} \\
-t_{AB} & 0 & 0 & \Delta + 2 \epsilon_a & -t_{AA} & 0 & 0 & t_{AB} & t_{AB} \\
0 & t_{AB} & 0 & -t_{AA} & \Delta + 2 \epsilon_a & 0 & t_{AB} & 0 & t_{AB} \\
0 & -t_{AB} & -t_{AA} & 0 & 0 & \Delta + 2 \epsilon_a & 0 & -t_{AB} & -t_{AB} \\
t_{AA} & -t_{AA} & -t_{AB} & 0 & t_{AB} & 0 & U_A + 2 \epsilon_a & 0 & 0 \\
t_{AA} & -t_{AA} & 0 & t_{AB} & 0 & -t_{AB} & 0 & U_A + 2 \epsilon_a & 0 \\
0 & 0 & -t_{AB} & t_{AB} & t_{AB} & -t_{AB} & 0 & 0 & U_B + 2 \Delta + 2 \epsilon_a
\end{pmatrix}
\end{equation}
The low-lying subspace consists of states where each A site is occupied by an electron, and all B sites are unoccupied. This subspace can be spanned by two basis states: \(c_{A_1\downarrow}^\dagger c_{A_2\uparrow}^\dagger\left.| 0 \right\rangle\) and \(c_{A_1\uparrow}^\dagger c_{A_2\downarrow}^\dagger\left.| 0 \right\rangle\).  We treat the off-diagonal terms outside this subspace as perturbations. Since there is no direct hopping between these two states, without perturbation, there is no spin-spin interaction favoring either larger or smaller total spin. However, when we introduce the interaction, terms contributing to the energy splitting appear.
\end{widetext}
\twocolumngrid

The perturbation procedure is as follows: First, we apply the standard perturbation formula, incorporating the effects of higher excited states while ensuring that the state remains orthogonal to the other original unperturbed state. Next, we use the Gram-Schmidt orthogonalization process on the state and use it to expand the original Hamiltonian, yielding an effective interaction induced by these higher-order states.  

The final reduced Hamiltonian has the non-diagonal term:
\begin{eqnarray}
    \mathcal{H}_{12}=\mathcal{H}_{21}&=&\frac{2t_{AA}^2}{U_A}  + \frac{2t_{AB}^2 (t_{AA} U_A + 2t_{AA} \Delta)}{U_A \Delta^2}-  \frac{8 t_{AA}^4}{U_A^3} \nonumber\\
    &&+ \frac{2t_{AB}^4(2U_A + U_B + 2\Delta)}{U_A \Delta^2 (U_B + 2\Delta)}%\label{effective exchange full}
\end{eqnarray}
For the Hamiltonian:
\begin{equation}
    \mathcal{H}=-J \mathbf{S}_1 \cdot \mathbf{S}_2
\end{equation}
The matrix representation in the basis \(c_{A_1\downarrow}^\dagger c_{A_2\uparrow}^\dagger\left.| 0 \right\rangle, c_{A_1\uparrow}^\dagger c_{A_2\downarrow}^\dagger\left.| 0 \right\rangle\) is:
\begin{equation}
    \mathcal{H}=J\begin{pmatrix}
        \frac14 & -\frac12\\
        -\frac12 &\frac14
    \end{pmatrix}
\end{equation}
Thus, the effective \(J\) is:
\begin{eqnarray}
    J &=&  - \frac{4t_{AA}^2}{U_A}  - \frac{4t_{AB}^2 (t_{AA} U_A + 2t_{AA} \Delta)}{U_A \Delta^2}+ \frac{16 t_{AA}^4}{U_A^3} \nonumber\\
    &&- \frac{4t_{AB}^4(2U_A + U_B + 2\Delta)}{U_A \Delta^2 (U_B + 2\Delta)}%\label{effective exchange full}
\end{eqnarray}
Upon inspecting the equation, we notice that since \(|t_{AA}|\ll U_A\), \(\frac{16 |t_{AA}|^4}{U_A^3}\ll \frac{4|t_{AA}|^2}{U_A}\). Thus, we can neglect the term \(\frac{16 |t_{AA}|^4}{U_A^3}\). Doing so reduces the equation to Eq.~\eqref{effective exchange full} in the main text. We then reorganize the equation to the following form:
\begin{equation}
    J =  \frac{t_{AA}^2}{U} \left( -4 + \frac{4A(1 + 2B)}{B^2} - \frac{4A^2(3 + 2B)}{B^2(1 + 2B)} \right)
\end{equation}
where $A=t_{AB}^2/(U (-t_{AA}))$ and $B=\Delta/U$. Here, we have set \(U_A=U_B=U\); if not, setting \(U_A=\lambda U_B=U\) only modifies some constants. The expression of \(J\) reveals competition between ferromagnetic and antiferromagnetic order at \(A, B\) both in order of one, making the region in specific interest. Moreover, the precise values of \(U\) and \(t_{AA}\) are not crucial; and even their ratio gets unimportant for determining the magnetic phase when \(U/|t_{AA}|\) grows large, as high order correction becomes negligible. 

Finally, we assert that no other terms provide corrections of the same or higher order. Our earlier perturbation analysis assumed \(t_{AA}, t_{AB}\) are in the same order and both much smaller than \(U, \Delta\). We ignored three-body terms of order \(t_{AA}^3\), but included superexchange terms of \(t_{AB}^4\). We now justify this by noting that \(|t_{AB}| \sim \sqrt{|t_{AA}| U}\) through nondimensionalization of the equations, so \(|t_{AB}|^4/(U^2 \Delta) \sim |t_{AA}|^2/\Delta \gg |t_{AA}|^3/U^2\). Consequently, these three-body terms can be safely regarded as higher-order corrections, consistent with our observation that at large \(U\), the phase boundary converges to the theoretically predicted result.

\section{\label{Calculation Detail for Real Hopping}Calculation Detail for Real Hopping}
We performed Exact Diagonalization (ED) and Density Matrix Renormalization Group (DMRG) calculations to validate our analytical results. The model Hamiltonian is given in equation~\eqref{complete hamiltonian}.

\subsection{ED calculation}
ED calculations were performed using the Quspin package \cite{weinbergQuSpinPythonPackage2019} for $3\times3$ and $3\times6$ lattices. We fixed $U_A=U_B=40 |t_{AA}|$ and scanned $t_{AB}$ and $\Delta$ for the phase diagram. 
\begin{figure}
    \centering
    \includegraphics[width=1\linewidth]{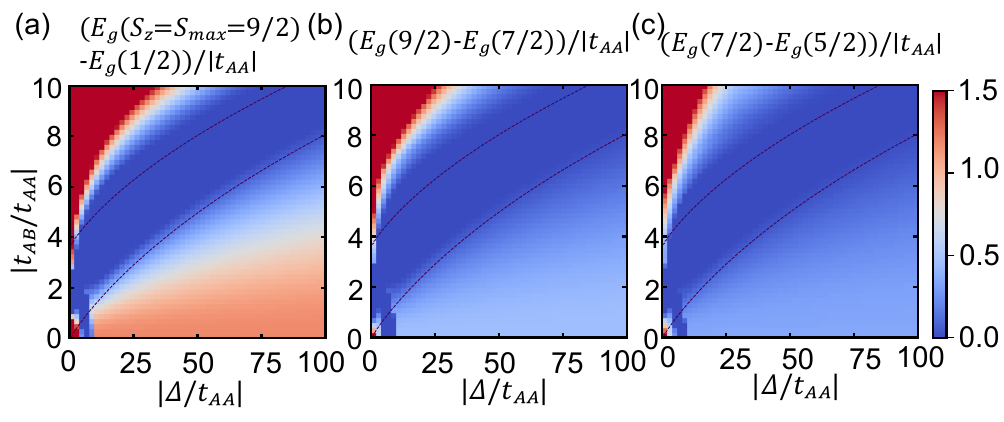}
    \caption{Spin gap results using different definitions. (a) Identical to Fig.~\ref{Fig2}(b) for reference, with the gap defined as the energy difference between $S_z=S_{max}=9/2$ and $S_z=1/2$. (b) Gap defined between $S_z=9/2$ and $S_z=7/2$. (c) Gap defined between $S_z=7/2$ and $S_z=5/2$. Dashed line here indicates phase boundary obtained from perturbation theory.}
    \label{FigS1}
\end{figure}
\begin{figure}
    \centering
    \includegraphics[width=1\linewidth]{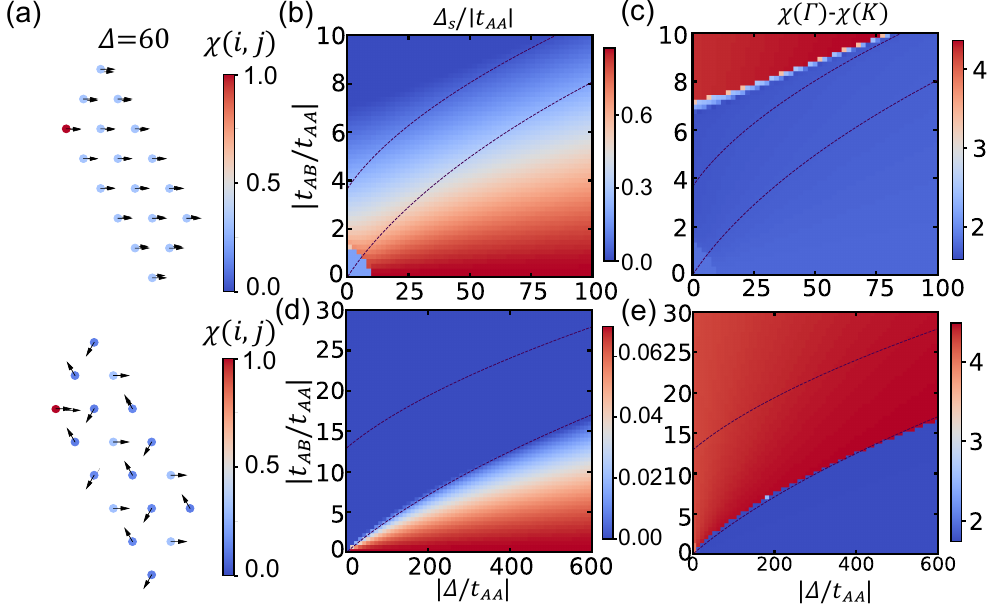}
    \caption{Results for $3\times6$ lattice with $U_A=U_B=40 |t_{AA}|$, $t_{AB}=10|t_{AA}|$, and spin sector $S_z=S_{max}-2$: (a) Real-space correlation. (b) The spin gap between $S_z=S_{max}$ and $S_z=S_{max}-2$. (c) Difference between spin-spin correlation at $\Gamma$ and $K$ points. (d,e) Same as (b,c) but with $U_A=U_B=500 |t_{AA}|$.}
    \label{FigS2}
\end{figure}
Since doing exact ED to a large system is impractical due to the exponential growth of the Hilbert space, we consider calculating the system with small lattice sites but without any cut-off, or consider calculating the system with large lattice sites and small cut-off. We first consider $3\times3$ lattice, in which through translation symmetry, rotation symmetry, and the conservation of particle number and spin, we can study it throughout without imposing any constraints. For the $3\times6$ lattice, we imposed constraints: maximum one particle on B sublattice and considering spin sectors from $S_z=S$ to $S_z=S-2$. As shown in Fig.~\ref{FigS1}, the spin sector constraints do not affect the identification of the phase boundary—the ferromagnetic phase boundaries from different pairs of spin sectors closely match those in the leftmost figure, where ferromagnetism is judged most strictly, requiring all \(S_z\) have the same lowest energy. The effect of particle constraint on B sublattice will be discussed later.

The results for \(3\times 3\) cluster are presented in the main text. The results of \(3\times 6\) cluster are shown in Fig.~\ref{FigS2}. In Fig.~\ref{FigS2}(b,c), we display the spin gap and correlation intensity. In the \(120^\circ\) AFM region, though exhibiting a significant increase of correlation at \(K\) point, the peak at the \(\Gamma\) point remains higher than that of the \(K\) point. We attribute this to the selected spin sector: \(S_z=S-2\), retaining some ferromagnetic characteristics due to the imposed constraints. Fig.~\ref{FigS2}(a) shows the real-space spin-spin correlations. At \(\Delta=200\), corresponding to the \(120^\circ\) AFM phase, spins that are expected to align exhibit stronger correlations compared to those that do not. Fig.~\ref{FigS2}(d,e) depict the same calculations as (b,c), but with \(U_A=U_B=500 |t_{AA}|\). As with the \(N_x=N_y=3\) case, increasing \(U\) leads to a phase boundary that more closely aligns with the proposed theoretical prediction.

While we observe a distinct phase boundary between the \(120^\circ\) AFM and FM phases, certain discrepancies remain. For example, the phase boundary for the AFM phase at small \(\Delta\) is missing, and the FM phase emerges at much larger \(t_{AB}\) than expected. We attribute these discrepancies to artifacts arising from the imposed cutoff. While the choice of \(n_B=1\) adequately captures the mechanics of ferromagnetism, as demonstrated in Fig.~\ref{Fig1}, it is insufficient to fully describe the AFM superexchange mechanism. Additionally, in the original case, though we only consider a few low order virtual process, we do not explicitly prevent two virtual processes from occurring simultaneously at different bounds. This resembles a random phase approximation, but under the current constraints, we effectively only account for the lowest-order diagrams due to the cut-off imposed. As a result, virtual processes at different bounds suppress each other, causing ferromagnetism to appear at larger \(t_{AB}\) and obscuring the existence of other phase boundaries. These observations suggest that even with an increased system size, compromises in the calculation persist, leading to results that may not be as accurate as expected. This limitation motivates our use of DMRG, which will circumvent these obstacles

\subsection{DMRG calculation}
DMRG calculations were performed using the ITensor package \cite{fishmanCodebaseReleaseITensor2022,fishmanITensorSoftwareLibrary2022}. We used a $3\times12$ lattice with periodic boundary conditions in the x-direction. Quantum numbers $(N_\uparrow, N_\downarrow)$ were conserved.

In calculation, we use a randomly generated matrix product state(MPS) with a fixed number of up and down spin as the initial state. The relative cutoff for the DMRG sweeping is set to \(1\times 10^{-10}\) and the absolute convergence criterion is \(5 \times 10^{-6}\). During the initial sweeps, we introduce noise with a magnitude of \(10^{-5}\), which is later reduced to \(10^{-7}\) and eventually to 0. To ensure convergence, we benchmark the results against those obtained with a stricter absolute convergence criterion of \(1 \times 10^{-6}\) and a cutoff of \(1\times 10^{-12}\), finding an energy difference on the order of \(0.001\), which is much smaller comparing to the spin gap, which is of the order of \(0.1\). This shows that relative good convergence has been reached. 

\begin{figure}
    \centering
    \includegraphics[width=1\linewidth]{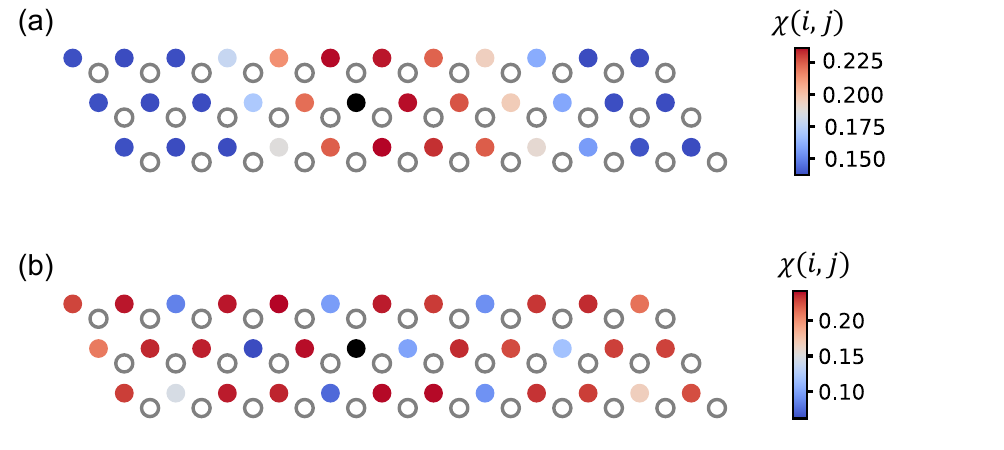}
    \caption{DMRG correlation results for $U_A=U_B=40 |t_{AA}|$: (a) $t_{AB}=10 |t_{AA}|$, $\Delta=100 |t_{AA}|$. (b) $t_{AB}=5 |t_{AA}|$, $\Delta=100 |t_{AA}|$.}
    \label{FigS3}
\end{figure}

Using DMRG we have calculated the spin gap and the spin-spin correlation in the sector of two spin flips. The spin gap is shown in the main text, and the correlation is shown in Fig.~\ref{FigS3}. We find agreement with the previous ED result that in the ferromagnetic phase, there will be a uniform polarization and in the antiferromagnetic phase there will be a periodic pattern in \(\sqrt{3} \times \sqrt{3}\) unit cell, in accordance to what the \(120^\circ\) AFM phase predicted. 

However, it is important to note that DMRG is not well-suited for capturing long-range correlations, as it struggles with the required long-range entanglement. The spin-wave excitations lead to a gapless spectrum, making convergence challenging, as discussed in Ref.~\cite{hastingsAreaLawOnedimensional2007a}. In Fig.~\ref{FigS3}(b), although the \(\sqrt{3} \times \sqrt{3}\) pattern is visible, it deviates noticeably from the spin correlation pattern obtained in systems of size \(3\times 3\) and \(3 \times 6\), as discussed in Appendix \Ref{Extra Data for Spin Correlation}. This discrepancy can be attributed to several factors. First, in the chosen spin-flip sector, the enforced reduction in spin misalignment prevents the AFM configuration from fully exploiting its favorable energetics, particularly in the presence of three-body interactions. Additionally, achieving a translation-invariant state using DMRG, similar to that produced by ED, is difficult because it requires superimposing of several translated images of a many-body state with exponentially suppressed hopping elements between them. Such a highly entangled state, which provides an exponentially small energy gain over its non-invariant counterpart, is challenging to resolve through imaginary time evolution.   

\section{\label{Extra Data for Spin Correlation}Additional Spin Correlation Data}
\begin{figure}
    \centering
    \includegraphics[width=1\linewidth]{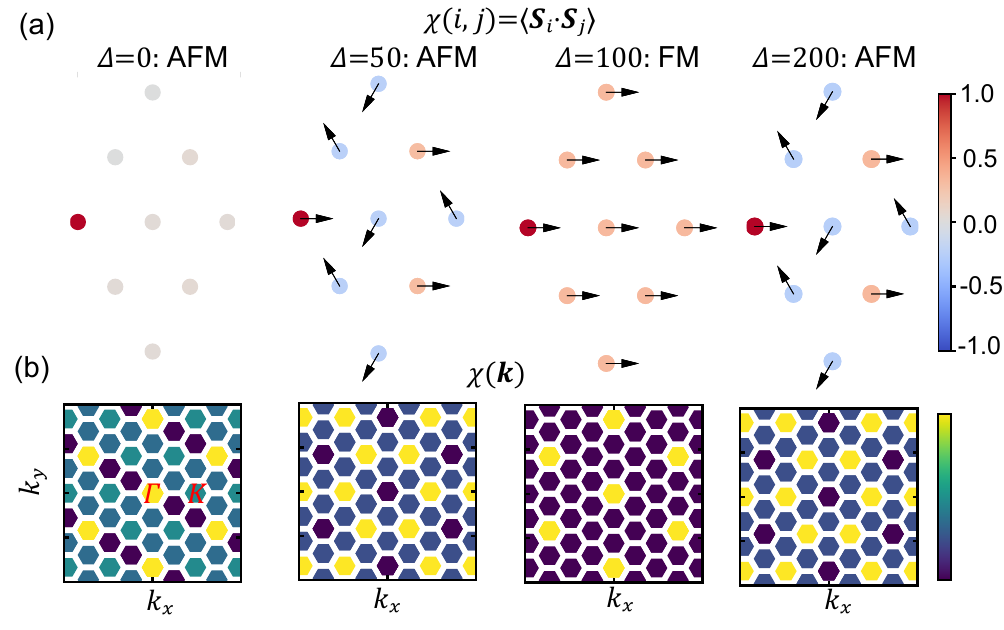}
    \caption{Spin-spin correlation for $U=40 |t_{AA}|$, $t_{AB}=10 |t_{AA}|$, and real $t_{AA}$. (a) Real-space correlation. Color indicates correlation strength relative to self-correlation. (b) Reciprocal space correlation. Color represents correlation intensity. $\Gamma$ and $K$ points are marked, where for simplicity we drop the scale. We have marked the corresponding point of \(\Gamma\) and \(K\) in the graph.}
    \label{FigS4}
\end{figure}

\begin{figure}
    \centering
    \includegraphics[width=1\linewidth]{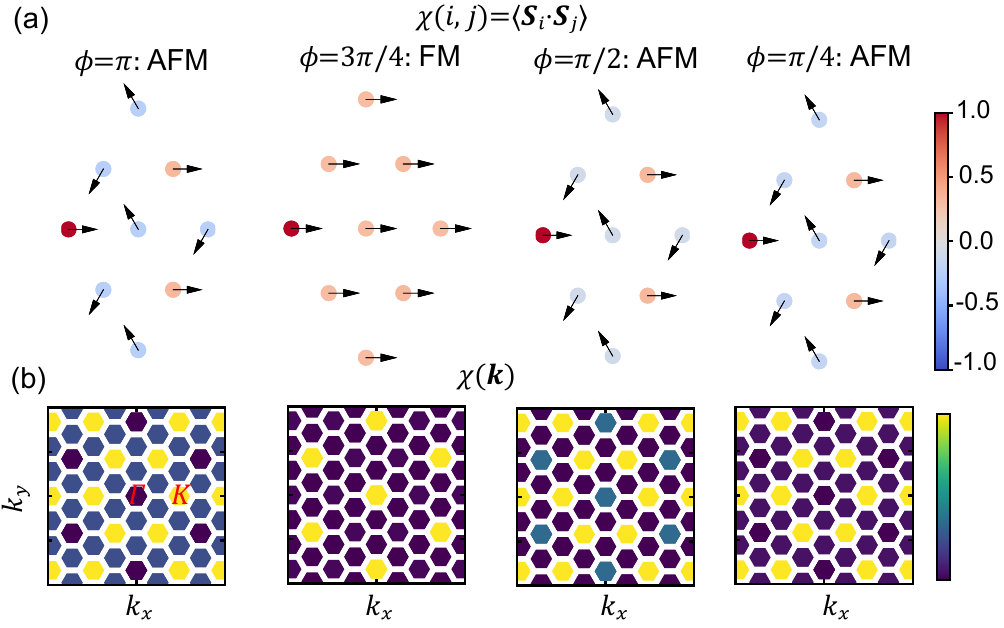}
    \caption{Spin-spin correlation for $U=40 |t_{AA}|$, $t_{AB}= 10 |t_{AA}|$, $t_{AA}=|t_{AA}| e^{i \phi}$, and $\Delta=150 |t_{AA}|$. (a) Real-space correlation. (b) Reciprocal space correlation. Color schemes are consistent with Fig.~\ref{FigS4}.}
    \label{FigS5}
\end{figure}

We present spin-spin correlation results from ED at various parameters, providing a detailed view of magnetic phases in different parameter regimes. Fig.~\ref{FigS4} shows results for real hopping, while Fig.~\ref{FigS5} presents data for complex hopping.

In Fig.~\ref{FigS4}(b), the ferromagnetic phase shows a prominent peak at the $\Gamma$ point, while the expected $120^\circ$ AFM phase exhibits a larger peak at the $K$ point, consistent with our previous discussion. The $\Gamma$ peak corresponds to positive nearest-neighbor correlations, while the $K$ peak indicates negative correlations, characterizing the $120^\circ$ AFM state (Fig.~\ref{FigS4}(a)). Regions with insignificant spin-spin correlation are also observed, potentially indicating paramagnetic or antiparamagnetic phases. Further investigation with larger system sizes and higher-order expansions is needed to confirm these phases.

Fig.~\ref{FigS5} shows similar patterns to Fig.~\ref{FigS4} for each phase type. The ferromagnetic phase boundary for complex hopping is discussed in Appendix~\ref{Effective spin model for complex hopping}.


\section{\label{Effective spin model for complex hopping}Effective Spin Model for Complex Hopping}
When setting $t_{AA}$ as a complex variable, the $\mathrm{SU}(2)$ symmetry breaks to $\mathrm{U}(1)$ symmetry, with only $S_z$ conserved. This scenario requires the inclusion of anisotropic terms in the Heisenberg model and the Dzyaloshinsky-Moriya interaction, breaking spin rotation and inversion symmetry.

We start with a comprehensive Hamiltonian incorporating complex hopping:

\begin{align}
&\mathcal{H}=H_A+H_B+H_{AB}+U_A\sum_{i\in A}n_{i\uparrow}n_{i\downarrow}+U_B\sum_{i\in B}n_{i\uparrow}n_{i\downarrow}. \\
&H_{\alpha}=-\sum_{\langle i,j\rangle_{\alpha},\sigma} (t_{\alpha}e^{is_{\sigma}\nu_{ij}\phi_{\alpha}}c_{i\sigma}^{\dagger}c_{j\sigma}+\mathrm{H.c.})-\sum_{i\in\alpha}\frac{\tau_{\alpha}\Delta}{2}n_{i}, \\
&H_{AB}=-t_{AB}\sum_{\langle i,j\rangle,\sigma} c_{i\sigma}^{\dagger}c_{j\sigma}. %\label{complete hamiltonian}
\end{align}
Here, $\alpha\in\{A,B\}$, $\tau_{\alpha}=\pm1$ for $A$ and $B$ respectively, \(\sigma\) indicates spin with \(s_\sigma \in \{+, -\}\), and $\nu_{ij}=\pm1$ for right/left-turning paths. We set $t_A e^{i \phi_A}=t_B e^{i \phi_B}=t_{AA}$ as the complex amplitude of nearest neighbor hopping, $t_{AB}$ as the real amplitude of second nearest neighbor hopping, $U_A, U_B$ as the on-site Coulomb repulsion at A, B sites, and $\Delta$ as the potential difference between sublayers.

Using perturbation theory approach similar to the one mentioned previously, we derive the effective spin-spin interaction Hamiltonian:

\begin{equation}
    \mathcal{H}=- \sum_{\langle\langle i,j\rangle\rangle}\left(J \boldsymbol{S}_i \cdot \boldsymbol{S}_j+J^\prime S_{i,z} \cdot S_{j,z} + J^{\prime\prime} (\boldsymbol{S}_i \times \boldsymbol{S}_j)\cdot \hat{z}\right)\label{spin_spin_interaction_complex}
\end{equation}
Where:
\begin{eqnarray}
    J&=&\frac{4 \text{Im}(t_{AA})^2}{U_A} - \frac{4 \text{Re}(t_{AA})^2}{U_A} - \frac{4 \text{Re}(t_{AA}) t_{AB}^2}{\Delta^2} \nonumber\\
    &&- \frac{8 \text{Re}(t_{AA}) t_{AB}^2}{U_A \Delta} - \frac{4t_{AB}^4(2U_A + U_B + 2\Delta)}{U_A \Delta^2 (U_B + 2\Delta)} \label{heisenburg exchange}\\
    %- \frac{16 \text{Im}(t_{AA})^4}{U_A^3} + \frac{16 \text{Re}(t_{AA})^4}{U_A^3}\label{heisenburg exchange}\\
    J^\prime&=&- \frac{8 \text{Im}(t_{AA})^2}{U_A}\label{anistropic}\\
    %+ \frac{32 \text{Re}(t_{AA})^2 \text{Im}(t_{AA})^2}{U_A^3} + \frac{32 \text{Im}(t_{AA})^4}{U_A^3}\label{anistropic}\\
    J^{\prime\prime}&=&- \frac{8 \text{Re}(t_{AA}) \text{Im}(t_{AA})}{U_A}-\frac{4 \text{Im}(t_{AA}) t_{AB}^2}{\Delta^2} - \frac{8 \text{Im}(t_{AA}) t_{AB}^2}{U_A \Delta}  \label{dm}\nonumber\\
    %&&+ \frac{32 \text{Re}(t_{AA})^3 \text{Im}(t_{AA})}{U_A^3} + \frac{32 \text{Re}(t_{AA}) \text{Im}(t_{AA})^3}{U_A^3}\label{dm}    
\end{eqnarray}

Here, we only preserve the term to the dominant order. The first term $J$ represents the Heisenberg exchange, reducing to the previous result in Eq.~\eqref{effective exchange full} when $\text{Im}(t_{AA})=0$. The second term $J^\prime$ introduces anisotropy, resembling the Ising model. The third term $J^{\prime\prime}$ is the Dzyaloshinsky-Moriya interaction, breaking inversion symmetry and potentially leading to non-coplanar magnetic states. \cite{yangFirstprinciplesCalculationsDzyaloshinskii2023,wangSpinChiralityFluctuation2019, moriyaAnisotropicSuperexchangeInteraction1960, dzyaloshinskyThermodynamicTheoryWeak1958}.

The anisotropic term in Eq.~\eqref{anistropic} favors states with small $S_z$ products, suggesting that the ferromagnetic (FM) ground state likely has the smallest possible $S_z$. The Dzyaloshinsky-Moriya term in Eq.~\eqref{dm} differentiates between clockwise and anticlockwise $120^\circ$ AFM states, with the state with anticlockwise rotation of the spin from \(j\to i\) as a right turning path would have lower energy under positive \(\phi\).

The shift of the FM phase towards larger $\Delta$ can be understood as a result of opposite phases of hopping amplitude in different layers reducing the effective exchange. As the imaginary part of $t_{AA}$ increases, the exchange term in Eq.~\eqref{heisenburg exchange} is reduced. This is captured by triangular Hubbard model with complex hopping  where ferromagnetism is reached in \(2 \pi/3>|\phi|>\pi/3\) \cite{panBandTopologyHubbard2020}. The ferromagnetic term's strength is also reduced as $\text{Re}(t_{AA})$. Thus, the emergence of ferromagnetism is a combined effect of kinetic exchange, the FM process, and superexchange, which remains antiferromagnetic and constant in strength as it depends only on $t_{AB}$. This interplay defines the phase boundaries of the AFM region.

\section{\label{large U real hopping}Calculation of phase diagram under large on-site potential }
We extend our calculations to the regime of large $U$ and $\Delta$, setting $U_A=U_B=500 |t_{AA}|$ and scanning $\Delta$ values comparable to $U$. This regime, with a smaller $t/U$ ratio, is expected to align better to our theoretical predictions.

\begin{figure}
    \centering
    \includegraphics[width=1\linewidth]{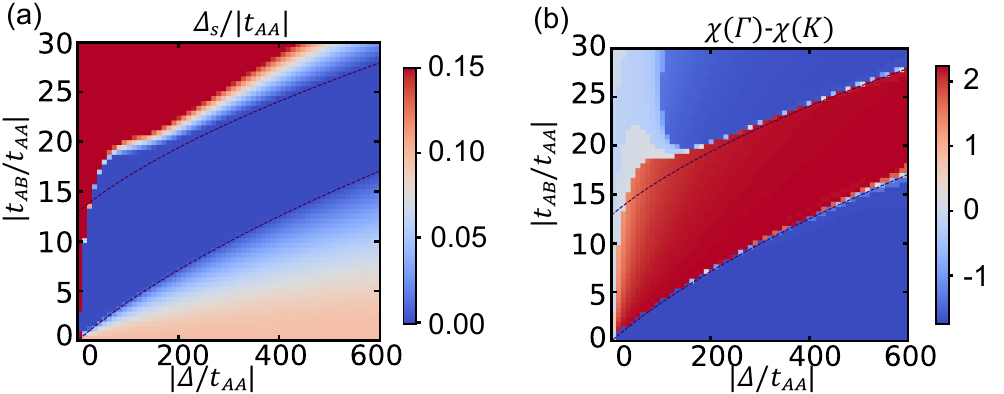}
    \caption{Results for $U_A=U_B=500 |t_{AA}|$ on a $3\times3$ PBC cluster using ED. (a) Spin gap $E_{g, Smax}-E_g$. A cutoff of 1.5 is imposed for clarity. (b) Correlation strength difference between $\Gamma$ and $K$ points.}
    \label{FigS6}
\end{figure}

Fig.~\ref{FigS6} presents the spin gap and correlation strength difference between $\Gamma$ and $K$ points. The results suggest three distinct phases. A ferromagnetic phase is observed, bounded by the theoretically proposed lines. On either side of this region, we find \(120^\circ\) AFM phases. Below $\Delta=100$, a potential non-magnetic state emerges, characterized by nearly equal peaks at $\Gamma$ and $K$ points, indicating negligible inter-site correlations.

The ferromagnetic region aligns closely with our perturbation analysis predictions. The emergence of the potentially non-magnetic state in the highly non-perturbative zone requires further investigation, involving higher-order expansions or alternative analytical approaches. These three phases constitute a comprehensive magnetic phase diagram for this parameter regime.

\section{\label{Connected to real material calculation}Connected to real material calculations}

Twisted bilayer transition metal dichalcogenides (TMDs) present exemplary systems for exploring the Hubbard physics, owing to their high tunability. These systems, composed by \(\mathrm{M X_2}\), (where M = Mo, W and X = S, Se, Te) exhibit electron localization primarily at \(\mathrm{XM}\) and \(\mathrm{MX}\) stackings. The stacking orientation between monolayers significantly influences electronic structures, with R-stacking breaking inversion symmetry and giving rise to exotic phases \cite{liRhombohedralstackedBilayerTransition2023}. Additionally, vertical electric fields can induce energy differences between localized positions, offering further tuning. From now on we separate our discussion into two cases: When moir\'e  minibands originating from the \(\Gamma\) point of the untwisted structure, or originating from the \(K\) point

Moir\'e bands originating from the \(\Gamma\) point of the untwisted structure, as identified in systems such as \(\mathrm{MoS_2/MoS_2}\) bilayers, exhibit approximated spin degeneracy due to \(C_{2y}\) symmetry \cite{zhangElectronicStructuresCharge2021}. This results in a nearly real hopping parameter \(t_{AA}\) in the Kane-Mele model. Interlayer tunneling \(t_{AB}\) is estimated to be of the order \(\sqrt{t_{AA} U}\). Further calculation through maximal Wannier center construction \cite{xuMaximallyLocalizedWannier2024} suggest the possibility of a small negative \(t_{AA}\). Though as a rough estimate, it shows that our mechanics proposed is within the parameter region of experimental interest.% calculation data is needed

In contrast, the moir\'e band originates from the untwist band at \(K\) point display complex \(t_{AA}\) through strong spin-orbit coupling. Though much more complicated, we speculate that recent experimental observation of the magnetism in the intermediate gating field is driven by our mechanics. 
%In the experiment, it is observed that both at filling \(\nu=1\) and \(\nu=2/3\)
In the experiment, it is observed that at filling \(\nu=1\)
the system is ferromagnetic in the intermediate gating field whereas in both large and small gating field AFM phase is observed, ruling out Nagaoka ferromagnetism for possible small doping or simple Hund's coupling explanation \cite{zengThermodynamicEvidenceFractional2023}. As for the lattice model, previous calculation \cite{xuMaximallyLocalizedWannier2024} show that the maximally localized Wannier centers for such a system are largely polarized in the top and bottom lattices in real space, around \(\mathrm{XM}\) and \(\mathrm{MX}\) stacking, respectively. It also reveals that \(t_{AA}\) will exhibit a complex phase around \(2 \pi/3\), which is also identified at other systems with moir\'e band of the same origin\cite{crepelAnomalousHallMetal2023}. Such phase angle provides it a negative real part but also a unnegligible imaginary part. 
%. This, however, is accompanied by a large imaginary part, having the same amplitudes but with opposite signs for different lattices and spins. 

The appearance of such a complex part in second nearest neighbor hopping provides the need to consider the dependence of ferromagnetism concerning the phase angle of the second next-nearest neighbor hopping. This is captured by the complex model in Eq.~\eqref{spin_spin_interaction_complex}. The ferromagnetic contribution is described by the term \(- \frac{4 \text{Re}(t_{AA}) t_{AB}^2}{\Delta^2} - \frac{8 \text{Re}(t_{AA}) t_{AB}^2}{U_A \Delta}\) in Eq.~\eqref{heisenburg exchange}. Ferromagnetic correlation in such mechanics persists when \(\text{Re}(t_{AA})<0\). However, additional complexity is introduced since in such cases the rotation symmetry of \(\mathrm{SU}(2)\) breaks into \(\mathrm{U}(1)\). This makes different total spin orientation have different ground state energy, especially between the one which the spin pointing out of the spin and the one which the spin lies within the plane. As Ising type spin-spin interaction from Hund's coupling is always present in the system, the spin polarization will eventually be orientated to the out-of-plane direction.
%This requires the presence of on-site anisotropic term to maintain out-of-plane spin polarization since otherwise, every single spin will have a tendency to lie with the smallest \(S_z\), according to the time-reversal invariant nature of the corresponding single particle model. 

The proposed theory is applicable in regimes where the kinetic exchange is significant, contrasting with the direct exchange dominance in the flat band limit \cite{repellinFerromagnetismNarrowBands2020}. Numerical estimation shows that in the realistic parameter setup, kinetic exchange is comparably larger than direct exchange. Our mechanics provide a source for magnetism that is observable in the experiment.

\end{document}